# Ancient eclipses and long-term drifts in the Earth–Moon system

## M. N. Vahia[1,]*, Saurabh Singh[2], Amit Seta[3] and B. V. Subbarayappa[4]

[1]Tata Institute of Fundamental Research, Mumbai 400 005, India
[2]Department of Electronics Engineering, Indian School of Mines, Dhanbad, Jharkhand 826 004, India
[3]Centre for Excellence in Basic Science, University of Mumbai, Vidhyanagari Campus, Mumbai 400 098, India
[4]No. 31, Padmanabha Residency, BSK III Stage, Bangalore 560 085, India

We study anomalies in the Earth–Moon system using ancient eclipse data. We identify nine groups of anomalous eclipses between AD 400 and 1800 recorded in parts of India that should have completely missed the subcontinent according to NASA simulations (Espenak, F. and Meeus, J., NASA/TP 2006–214141, 2011). We show that the typical correction in lunar location required to reconcile the anomalous eclipses is relatively small and consistent with the fluctuations in the length of day that are observed in recent periods. We then study how the change in the moment of inertia of the Earth due to differential acceleration of land and water can account for this discrepancy. We show that 80% of these discrepancies occur when the Moon is at a declination greater than 10° and closer to its major standstill of 28° while it spends 46% of the time in this region. We simulate the differential interaction of the Moon's gravity with land mass and water using finite element method to account for land mass and watermass. We show that the results of eclipse error are consistent with the estimate of a small differential acceleration when the Moon is over land at high latitudes. However, we encounter some examples where the results from simulation studies cannot explain the phenomenon. Hence we propose that the ΔT corrections have to be coupled with some other mechanism, possibly a small vertical oscillation in the Moon's rotational plane with period of the order of a few hundred years to achieve the required adjustment in eclipse maps.

THE movement of the Moon and the differential changes in the gravitational field of the Earth are closely monitored using a slew of satellites and atomic clocks. These studies show that there is a small and apparently irregular fluctuation in the length of day that rides over a general seasonal variation[1]. The rotation of the Earth is also studied in significant detail[2] using multi polynomial expansion series which account for most of the changes in the Earth–Moon system over the last few hundred years.

However, the Earth–Moon system is far more complex with uneven distribution of mass and a fluid and solid component on the surface[3]. Apart from the Earth–Moon interaction, geological factors such as core–mantle coupling, mass redistribution due to glaciation and deglaciation, etc. also play a part in changing the length of day[4–9]. Lambeck showed that there is no cumulative change in the Earth–Moon system due to the tidal effect in the last 2000 or 3000 years. Similarly, Hansen[6] showed that the evolution of the orbit is not greatly influenced by a realistic tidal model. All these studies however, focus on long-period cumulative effects on the system.

The system is made more complex by the fact that the temperature fluctuations over a year have changed the mass distribution on Earth. Hence small fluctuations in the length of day are not fully amenable to an analytical solution even with more than 120 terms[10]. However, the fluctuations are minor and not fully exposed by short-period studies of a few decades. To understand the nature of interaction, it is important to have long-period data[11]. Based on individual eclipse records it has been shown that the drift time ΔT that characterizes long-term drift in the Earth–Moon system is not as smooth as conventionally thought (see refs 12–15 for a survey of the field). Here we use a database of more than 500 solar eclipses recorded in India between AD 400 and 1800 to quantify the drift in the location of the Moon with time.

## Present study

Subbarayappa and Vahia[16] have made a catalogue of more than a thousand eclipses recorded in the subcontinent from AD 400 to 1800. A small study of this nature was done by Shaylaja[17]. In the present study we compare the ancient Indian records of solar eclipses from AD 400 to AD 1500 (see also ref. 18). We find more than 500 solar eclipses in this period. Out of these, 114 are mentioned in more than one record. We then compare these observations with eclipse predictions by Espenak[19] (http://eclipse.gsfc.nasa.gov/eclipse.html). We find that 15 of these eclipses have paths that did not pass over the region at all where the observations are recorded. We then study single observations of eclipses around these

*For correspondence. (e-mail: vahia@tifr.res.in)

**Table 1.** List of 15 anomalous ancient eclipses whose paths as observed are in variance with Epstine

| Sl. no. | No. of observations* | Reference/source | Location | Date |
|---|---|---|---|---|
| 1.a | 2 | A.R.831 (1962; 142) | Tadkal, Manvi Tq., Raichur Dist., Karnataka | AD 1033, 4 Jan |
| 1.b | | KUES, Pt.5-2, 552 | Vadarapalli, Andhra Pradesh | AD 1033, 4 Jan |
| 2.a | 2 | KUES, Pt.6, 66 | Komarapalli, Deggur Tq., Nanded Tq., Maharashtra | AD 1079, 1 Jul |
| 2.b | | A.R.28 (1950; 14) | Prince of Wales Museum, Mumbai; find spot not known | AD 1079, 1 Jul |
| 3.a | 4 | A.R.527 (1914; 54) | Holagondi, Bellary Dist., Karnataka | AD 1083, 14 Oct |
| 3.b | | SII XX, 287 | Jamakhandi, Bijapur Dist., Karnataka | AD 1083, 14 Oct |
| 3.c | | EI XXXI, 33 | Kadonal, Udaipur Dist., Rajasthan | AD 1083, 14 Oct |
| 3.d | | A.R.215 (1946; 38) | Saligundi, Koppal, Hangal Tq., Dharwad Dist., Karnataka | AD 1083, 14 Oct |
| 4.a | 4 | EC IV, Hs.35 | Kiranguru, Hunsuru Tq., Mysore Dist., Karnataka | AD 1091, 21 May |
| 4.b | | A.R.8 (1974; 14; copper) | Rajpur, West Nimar Dist., Madhya Pradesh | AD 1091, 21 May |
| 4.c | | A.R.69 (1961; 50) | Bhuvanagiri, Nalagonda Dist., Andhra Pradesh | AD 1091, 21 May |
| 4.d | | A.R.128 (1913; 15) | Huvinahadagalli, Bellary Dist., Karnataka | AD 1091, 21 May |
| 5.a | 2 | A.R.123 (1959; 57) | Local Museum, Alampur Tq., Mahabubnagar Dist., Andhra Pradesh | AD 1097, 16 Jan |
| 5.b | | A.R.22 (1960; 40; copper) | Sivakasi, Sattur Tq., Ramanathapuram Dist., Tamil Nadu | AD 1097, 16 Jan |
| 6.a | 2 | A.R.316 (1915; 28) | Tangodumalle, Narasaraopetta Tq., Guntur Dist., Andhra Pradesh | AD 1115, 23 Jul |
| 6.b | | SII XVIII, 113 | Agadi (Haveri), Karnataka | AD 1115, 23 Jul |
| 7.a | 3 | A.R.70 (1941; 137) | Rachanapalle, Anantapur Dist., Andhra Pradesh | AD 1124, 11 Aug |
| 7.b | | A.R.452 (1959; 93) | Chitapur Tq., Gulbarga Dist., Karnataka | AD 1124, 11 Aug |
| 7.c | | A.R.553 (1958; 80) | Hirekerur Tq., Dharwar Dist., Karnataka | AD 1124, 11 Aug |
| 8.a | 6 | SII, IX-pt.II 200 | Kurtakoti, Godag Tq., Dharwad Dist., Karnataka | AD 1126, 22 Jun |
| 8.b | | SII IX-pt.I, 210 | Holalagundi, Alur Tq., Bellary Dist., Karnataka | AD 1126, 22 Jun |
| 8.c | | SII IX-pt.I, 211 | Bagali, Harapanahalli Tq., Bellary Dist., Karnataka | AD 1126, 22 Jun |
| 8.d | | SII IX-pt.I, 212 | Tripurantakam, Markapur Tq., Kurnool Dist., Andhra Pradesh | AD 1126, 22 Jun |
| 8.e | | A.R.638 (1958; 89) | Tadpillivarihill, Shiralli Tq., Dharwad Dist., Karnataka | AD 1126, 22 Jun |
| 8.f | | SII IX-pt.I, 123 | Bagali, Harapanahalli Tq., Bellary Dist., Karnataka | AD 1126, 22 Jun |
| 9.a | 3 | SII XX, 104 | Malghan, Sindagi Tq., Bijapur Dist., Karnataka | AD 1133, 2 Aug |
| 9.b | | A.R.8 (1956; 6; copper) | Paragoan, Raipur Dist., Madhya Pradesh | AD 1133, 2 Aug |
| 9.c | | A.R.16 (1956; 40) | Undavalli, Guntur Dist., Andhra Pradesh | AD 1133, 2 Aug |
| 10.a | 4 | EI XXIX, 27 | Bilaigarh, Raipur Dist., Madhya Pradesh | AD 1144, 26 Dec |
| 10.b | | SII XX, 112 | Sambi, Kundagol Tq., Dharwad Dist., Karnataka | AD 1144, 26 Dec |
| 10.c | | SII XV, 28 | Kodikop, Ron Tq., Dharwad Dist., Karnataka | AD 1144, 26 Dec |
| 10.d | | A.R.231 (1953; 44) | Haliyal Tq, North Carana Dist., Karnataka | AD 1144, 26 Dec |
| 11.a | 2 | A.R.530 (1914; 54) | Uttangi, Hadagali Tq., Bellary Dist., Karnataka | AD 1155, 26 Nov |
| 11.b | | A.R.344 (1937; 42) | Madala, Sattenapalle Tq., Guntur Dist., Andhra Pradesh | AD 1155, 26 Nov |
| 12.a | 3 | EI XXVIII, 40 | Nagari, Cirttack Dist., Odisha | AD 1230, 14 May |
| 12.b | | IKW, 74 | Motupalli, Bapatla Tq., Guntur Dist., Andhra Pradesh | AD 1230, 14 May |
| 12.c | | EC XII, TK.78 | Mudigere, Tarikere Tq., Chikkamagulur Dist., Karnataka | AD 1230, 14 May |
| 13.a | 2 | A.R.488 (1914; 49) | Hyarada, Hadagalli Tq., Bellary Dist., Karnataka | AD 1261, 1 Apr |
| 13.b | | A.R.489 (1914; 49) | Kattebennur, Hadagali Tq., Bellary Dist., Karnataka | AD 1261, 1 Apr |
| 14.a | 2 | SII XXVII, 185 | Hamdadi, Udupi Tq. and Dist., Karnataka | AD 1542, 11 Aug |
| 14.b | | EC X, (old ed.) Tl.162 | Mukkadegutte, Malur Tq., Kolar Dist., Karnataka | AD 1542, 11 Aug |
| 15.a | 2 | EC X, (old ed.) Mr.57 | Masti, Malur Tq., Kolar Dist., Karnataka | AD 1578, 8 Mar |
| 15.b | | A.R.56 (1931; 12) | Mukdukulattur Tq., Ramnad Dist., Tamil Nadu | AD 1578, 8 Mar |

*A.R., Annual Reports on Epigraphy of the Archaeological Survey of India; EC, *Epigraphia Carnatica*; IKW, Inscriptions of Kaokataya of Warangal; SII, South Indian Inscriptions; KUES, *Karnataka University Epigraphic Series*; EI, *Epigraphia Indica*.

periods and we show that there is a general trend of several closeby eclipses whose paths also do not fall as expected. In order to determine the time interval of anomalous eclipses, we assume that an eclipse is anomalous if the disc coverage at the location of observation is less than 20%, as predicted by Espenak[19]. We extend the period on either side of the observations till we come across two consecutive observations where the expected and observed paths agree. We find nine distinct periods when eclipses around these multiple observations also have paths which do not pass through the point of observation. We label them as anomalous eclipse periods. These lie in the range given in Table 1.

This anomaly may be caused by subtle differences in the Moon's location compared to the values used in the NASA calculations[20]. We then study the location of the Moon during these anomalous periods and find that it is significantly closer to major standstill during this period. We then calculate the differential acceleration of the Earth due to its varying oblateness because of the fact that water mass has a much greater displacement compared to land mass due to lunar gravity[21]. We show that this difference in moment of inertia of the Earth is consistent with fluctuations in the length of day and the errors in location of the Moon derived from our calculations. This is consistent with the fluctuations observed in

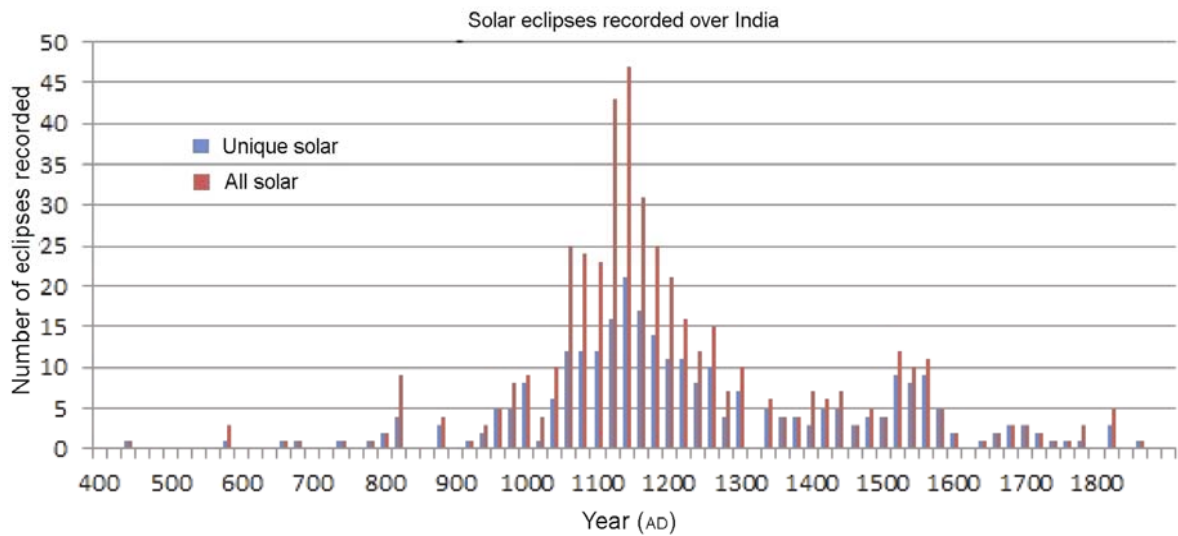

**Figure 1.** Distribution of eclipse records in ancient India.

**Table 2.** List of nine sets of anomalous eclipses

| Period | Calendar period | Interval | Number of eclipses | No. of anomalous eclipses (% of total eclipses) | Period width | Mean year | Difference from previous mean year | Difference as multiple of 9.3 years |
|---|---|---|---|---|---|---|---|---|
| 1 | 1007 to 1046 | 39 | 14 | 10 (71%) | 19.5 | 1026.5 | | |
| 2 | 1068 to 1108 | 60 | 17 | 7 (41) | 20 | 1088 | 61.5 | 6.6 |
| 3 | 1109 to 1122 | 13 | 7 | 3 (43%) | 6.5 | 1115.5 | 27.5 | 3.0 |
| 4 | 1123 to 1134 | 11 | 8 | 6 (75%) | 5.5 | 1128.5 | 13 | 1.4 |
| 5 | 1135 to 1163 | 28 | 13 | 8 (62%) | 14 | 1149 | 20.5 | 2.2 |
| 6 | 1228 to 1239 | 11 | 6 | 3 (50%) | 5.5 | 1233.5 | 84.5 | 9.1 |
| 7 | 1250 to 1290 | 40 | 8 | 4 (50%) | 20 | 1270 | 36.5 | 3.9 |
| 8 | 1527 to 1550 | 23 | 9 | 5 (56%) | 11.5 | 1538.5 | 268.5 | 28.9 |
| 9 | 1567 to 1590 | 23 | 7 | 3 (43%) | 11.5 | 1578.5 | 40 | 4.3 |

East Asian records[15] and the fluctuations in length of day recorded since the advent of atomic clocks.

## Eclipse data

Solar eclipses constrain the location of the Sun and Moon with significant accuracy and hence records of ancient eclipses have been used to determine the difference in the location of the Sun and the Moon[22]. We have catalogued a total of 529 records of solar eclipses between AD 400 and 1800 (ref. 16). These are largely recorded in stone inscriptions or copper plates and are listed in the *Inscriptionum Indiacanum* and other records (Figure 1).

The Indian records are based on donations granted on account of the observation of the eclipse. Hence it is safe to assume that they are not predictions and are records of direct observations against which grants were made. The dating of the eclipse has been through recalculation of the Indian calendrical dates (based on lunar mansions) con-verted to the Julian calendar. If the error is of a few days in the calculation (less than 1 month), then we assume that there is an error in recording and that the date as given by NASA calculations is the correct date, even if the path does not match accurately. In all the cases presented here, the NASA dates and inscription dates agree within this rider. In 80% of the cases, the NASA dates and Indian converted dates agree. Also, considering that we have identified the anomalous eclipses based on multiple observations from two or more separate regions, we believe that the identification is accurate. As it is rare even to have two eclipses in a year over the country, this is a good assumption.

In order to avoid possible errors or false records of eclipses, we studied 114 eclipses that were recorded at more than one location that is separated by several tens of kilometres. The records do not mention the fraction of the Sun's disc covered by the eclipse or the time at which they were observed. Hence we are unable to do a detailed calculation of the type done with East Asian records[13].

We, however, compare the records of eclipses with those of NASA[19]. All the eclipses in our database are found in Espenak[19]. We find 15 eclipses (Table 1) that, according to Espenak[19], should not have been visible where they were recorded. In Figure 2 we give the example of one such eclipse path. We investigate these eclipses in detail by studying some around that period. We find nine time intervals of eclipses which need special investigation (Table 2). The width of the period is decided when two consecutive eclipses around the anomalous eclipse are such that they would have been visible at 50% obscuration of the solar disk.

We calculate the maximum vertical and horizontal shift required for the eclipse to be visible at the recorded place of observation. In order to do this, we assume that the eclipse is considered as recorded if the eclipse phase is 50% or more at the location of the eclipse. In Figure 3, we have plotted the maximum tolerance for eclipses recorded between AD 1010 and 1045. As can be seen from the figure, several eclipses require significant corrections in the vertical location of the Moon for them to be visible at the location of observation.

We have done similar calculations since some eclipses could also be visible if the Moon was faster or slower than the value used in NASA calculations. We find that no significant changes in the time-period are required.

## Simulation studies

In order to study the possibility of unaccounted fluctuations in the lunar motion, we have simulated the Earth–Moon interaction using finite element method. We divide the globe of the Earth into about 42,000 blocks of size 111 km corresponding to $1° \times 1°$ at the equator and calculate the attraction of the Moon and resultant displacement or additional force at each location. Details of the formulation are given in the Appendices 1 and 2. Based on this we calculate the changes in the moment of inertia of the Earth due to increased oblateness. We assume each block to be about 2 km thick and the maximum displacement of water due to the lunar attraction to be 5 m. We assume that the land mass density is 10,000 kg/m$^3$ against 1000 kg/m$^3$ for water. For the ease of simulation, we divide the land mass of the Earth into blocks (Figure 4). The moment of inertia calculated using the finite element method was compared with the equation for moment of inertia for a spherical shell and the simulation agreed with the spherical shell with accuracy better than 0.02%. For the purpose of simulation, the Earth therefore looks as shown in Figure 5.

We then calculate the attractive force of the Moon experienced by the Earth when the Moon is at different locations. The attractive force will depend on the latitude and longitude of the Moon. In Figures 6 and 7 we show the attractive force when the Moon is at 0° long. and 28°N. The two curves indicate the nature of the force without the land mass at 0° lat. and when land mass is added and the Moon is at 28°N lat. Note that the force increases significantly over the land mass (which is assumed not to be displaced). Figure 7 shows the force experienced by the Moon when it is at different longitudes for two different latitudes.

In Figure 8 we show the simulation of oceanic tides due to the Moon when it is over the International Date Line (Pacific Ocean) and at 0°. As can be seen from the figure, the displacement of Earth mass is maximum at the region directly between the centre of the Earth and the Moon and on the opposite side[23].

The simulation gives an accurate measure of the differential force experienced by the Moon when it is over land and over water even though the magnitude of difference is of the order of 10$^{-8}$. We therefore use these simulations to calculate the moment of inertia (Figure 9) of the Earth when the Moon is over equator and when it is at 28°N using finite element method[24]. As can be seen from the figure, when the Moon is over the equator, the moment of inertia is maximum. Also, because there is significantly higher land mass in the northern latitudes, there is a crossover in the figure with higher angular momentum when the Moon is in the south compared to when it is at the equator.

We apply inverse relation between $I$ and $\omega$ of the Earth, where $I$ is the moment of inertia and $\omega$ is the angular velocity[6,25,26]. Hence we correlate the change in the length of day (LOD) when the Moon is at different latitudes (Figure 10).

LOD values used in the graph are the excess of those over 86400s. As can be seen from the figure, the value of LOD is maximum when the Moon is near the equator. This is due to maximum bulge at the equator so that $I$ is maximum and $\omega$ is minimum. Therefore, the Earth's rotation is the slowest and hence the LOD value is highest.

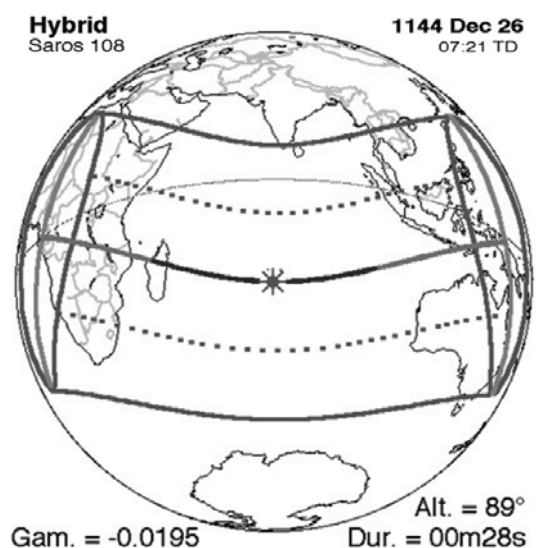

**Figure 2.** Path of eclipse of AD 1144. Eclipse map courtesy Fred Espenak, NASA/Goddard Space Flight Center, USA.

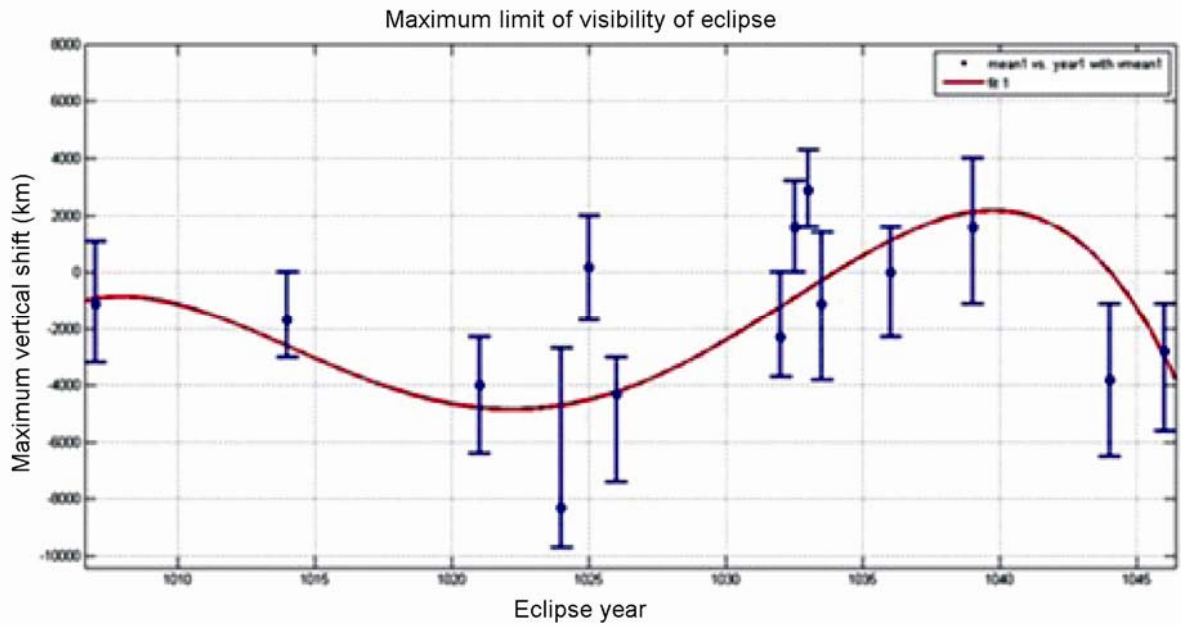

**Figure 3.** Maximum limit of visibility of eclipse. Zero value indicates that the eclipse would be visible at 50% or higher level at the location of observations. Note that for several eclipses the correction required in the location of the Moon is significant.

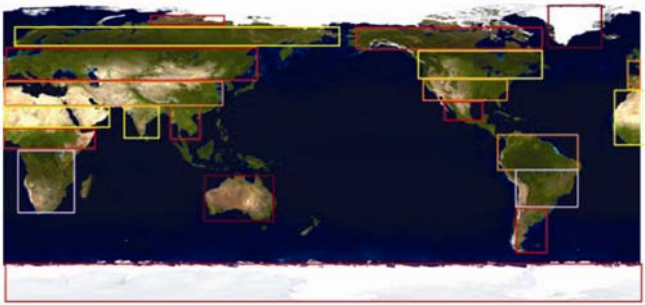

**Figure 4.** Division of the Earth into small blocks for simulation.

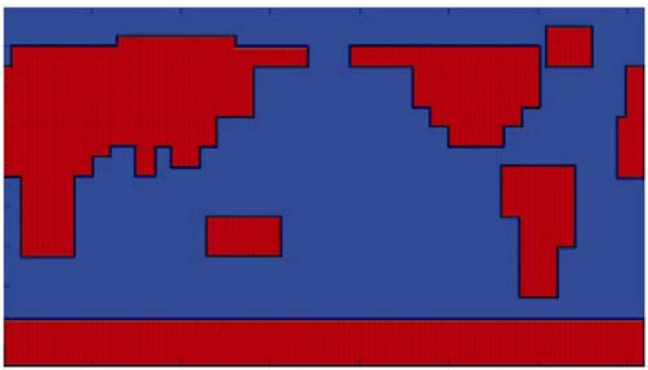

**Figure 5.** Map of the Earth used in the simulation.

We confirm this by comparing the change in LOD observed in June 2006 (ref. 27) and note that both the periods of negative changes (on 12 June and 27 June) correspond to the Moon being at the southern standstill at (28°S) and at upper standstill (28°N) respectively. Note that the change is more significant when the Moon is at 28°S (Figure 11).

## Perturbations in Moon's motion due to Earth–Moon interaction

We measured the minimum $\Delta T$ corrections for anomalous eclipses to be visible in India. However, we find that majority of instances require huge $\Delta T$ corrections that are not possible keeping in view the compatibility of world data. Hence we study the effect of changes in moment of inertia of the Earth on the moon as a possible reason for the anomalous eclipses.

We consider the Earth–Moon as an isolated system and hence conserve its total angular momentum.

$$H_0 = I\omega + M'nr^2,$$

where $I$ is the moment of inertia of the Earth, $\omega$ the angular velocity of the Earth, $M'$ the reduced mass of the system

$$= \frac{M_0 M_e}{M_e + M_0},$$

$M_e$ is the mass of the Earth and $M_0$ the mass of the Moon, $n$ the angular velocity of the Moon and $r$ is the distance of separation of Earth and Moon. We neglect the rotational

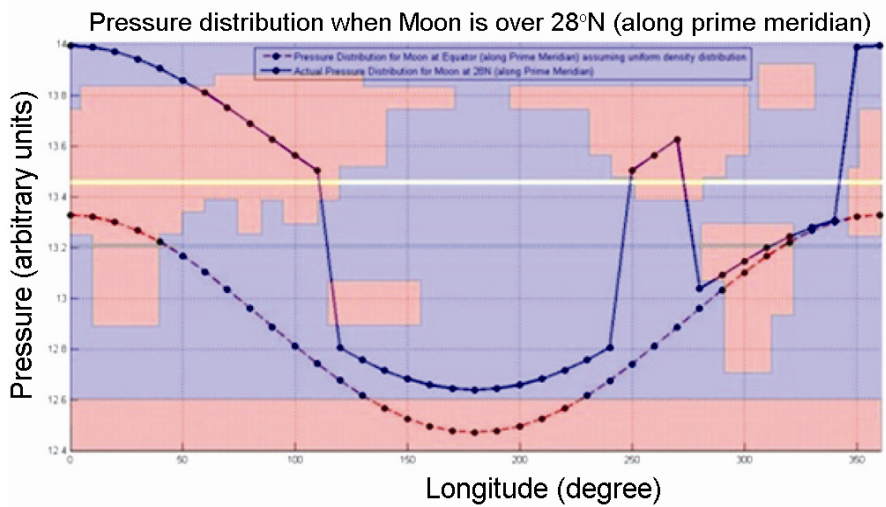

**Figure 6.** The pressure experienced by the Earth when the Moon is at 0° long. and has no land mass, and at 28°N lat. with the Earth land mass approximated according to Figure 5.

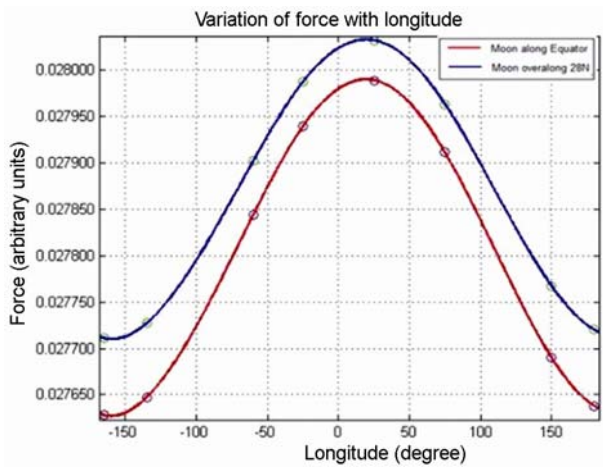

**Figure 7.** Calculation of the force experienced by the Moon at different longitudes.

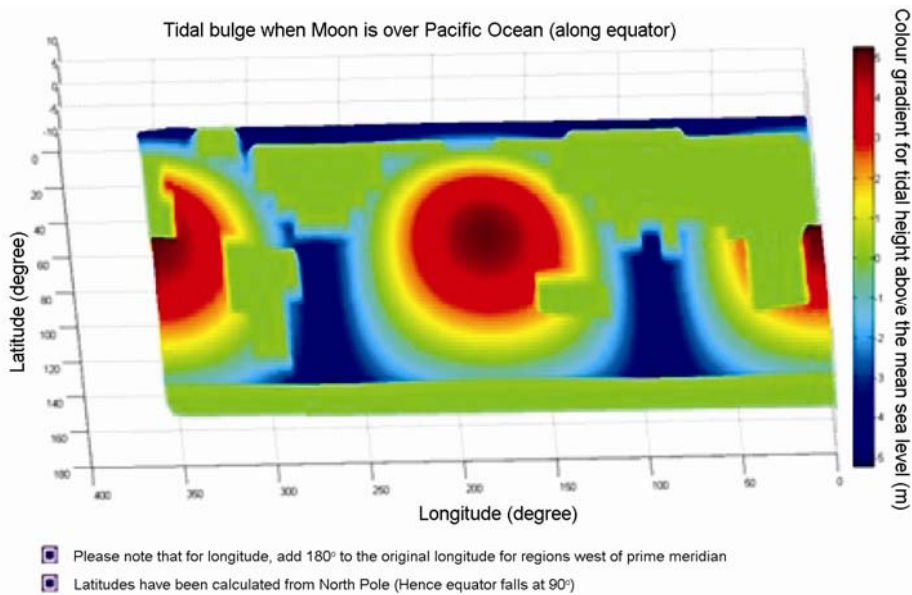

■ Please note that for longitude, add 180° to the original longitude for regions west of prime meridian
■ Latitudes have been calculated from North Pole (Hence equator falls at 90°)

**Figure 8.** Calculation of the tides (maximum level 5 m) when the Moon is over 180°E and equator.

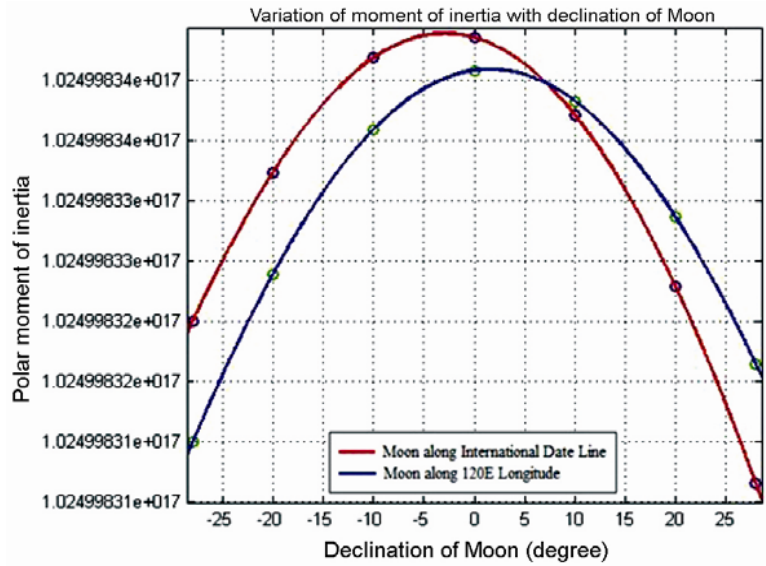

**Figure 9.** Change in the moment of inertia of the Earth as a function of the declination of the Moon, when the latter is over International Date Line and over 120°E long. when there is maximum land mass on either side.

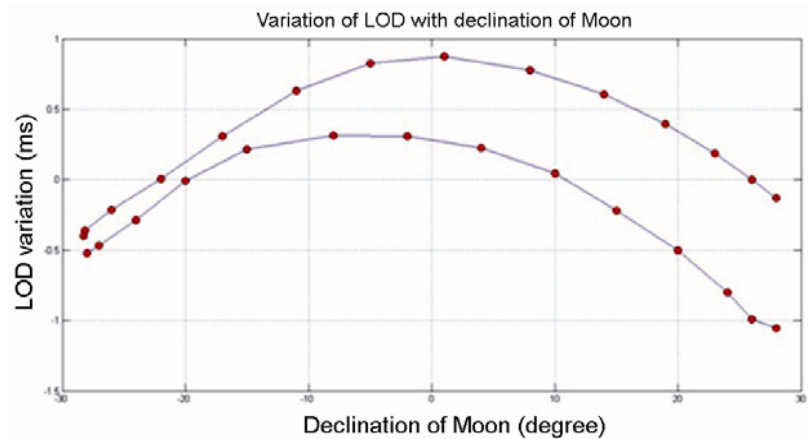

**Figure 10.** Change in length of day (LOD) when the Moon moved from lower standstill (–28°) to upper standstill (28°) in 2005 and 2006. The reason for the difference in the Earth's response is due to the different trajectories of the Moon over the Earth.

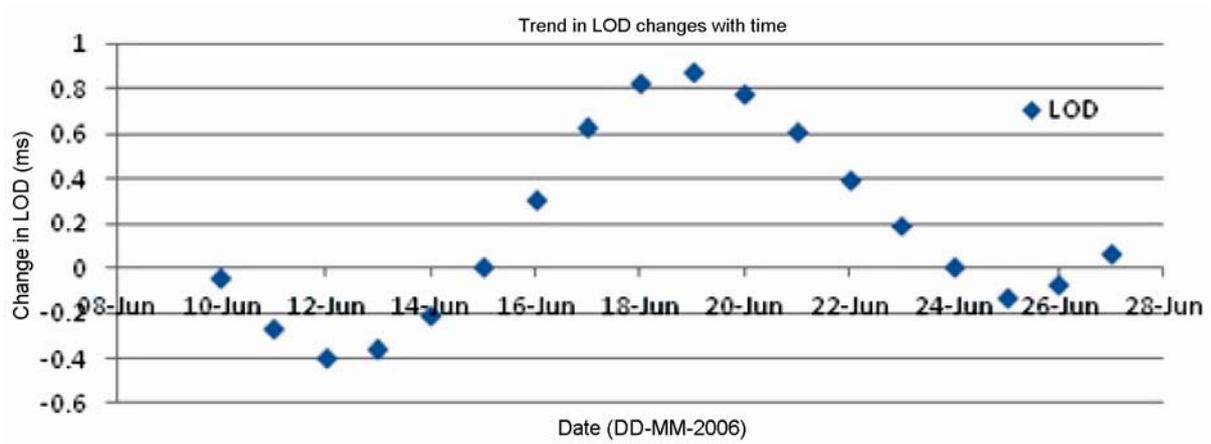

**Figure 11.** Change in LOD between 8 and 28 June 2006. Note that the two minima arise when the Moon is at 28°S and 28°N respectively.

angular momentum of the Moon in the above expression because the term is of the order of $10^{-8}$, which is too small compared with the other terms.

Since $H_0$ is a constant quantity, we try to measure change in angular velocity of the Moon:

$$\mathrm{d}n = -\frac{\omega\,\mathrm{d}I + I\,\mathrm{d}\omega}{M'r^2}. \qquad (1)$$

Measuring the order of this effect, we check the orders of magnitude of the involved quantities:

- $\omega \sim 10^{-5}$ rad/s
- $I \sim 10^{37}$ kg m$^2$
- $\mathrm{d}I \sim 10^{-9} \times 10^{37}$ kg m$^2$
- $\mathrm{d}\omega \sim 0.843994809 \times 10^{-12}*$ (change in length of day duration) rad/s
- $\sim 10^{-16}$ rad/s
- $M' \sim 10^{22}$ kg
- $r \sim 10^{8}$ m.

The value of $\mathrm{d}\omega$ is taken from Aoki *et al.*[28]. Hence change in orbital velocity of the Moon ($n \times r$) is $\sim 10^{-7}$ m/s.

Though the change is small, under extreme conditions (e.g. high declination of the Moon, perigee), it can cause slight perturbation in orbital velocity that can result in shift of the shadow during eclipses either in latitudes or longitudes (depending on the exact geometry of the situation). There are certain cases possible where the Moon can either slow down or speed up as a result of the above interaction and hence can shift the shadows of Moon on Earth in a dramatic manner.

The exact nature of the interaction will vary depending on whether the Moon is (a) in the ascending node and northern hemisphere, (b) descending node in the northern hemisphere, (c) ascending mode in the southern hemisphere or (d) descending mode in the southern hemisphere. In view of the uneven distribution of the land mass in the north and the south, the vector force can change the velocity or give an additional orthogonal acceleration. Hence the exact impact of these parameters will depend on the precise location of the Moon at each individual occasion. In general, the effect on the movement of the Moon can be in the form of slowing down or a vertical displacement in the Moon's orbit by an amount of a few kilometres resulting in significant shift in the location and appearance of the eclipse on the Earth.

## Comparison of Moon's location and ancient eclipses

In Figure 12 we have presented the $\Delta T$ values for fitting the eclipses compared with the world data. The method of calculating $\Delta T$ is as used by Soma and Tanikawa[12]. There were several possible cases in the analysis of anomalous records:

(1) $\Delta T$ correction was not at all possible in 13 cases.
(2) $\Delta T$ correction was possible in 32 cases, but a longitudinal shift was required. In these cases, there was no control on magnitude by $\Delta T$ correction since the shadow was moving only sideways. The magnitude of eclipses obtained after correction was from total to 0.1.
(3) $\Delta T$ correction was possible along with a latitudinal shift in 11 cases. These were the records where the horizontal movement of shadows through $\Delta T$ corrections resulted in slight vertical movement of shadows to reach India. However, in spite of the corrections, the eclipse was found to have very less visibility (around 0.1–0.2).

Based on this, we study the periods of the eclipses that were observed as predicted and those that were not observed. In Figures 13 and 14 we plot the longitude of the Moon during confirmed eclipses and during eclipses which should not have been observed. In Figure 15, we have plotted the location of the Moon during normal and anomalous eclipses. As can be seen from these figures, the discrepancy in the observation of eclipses arises primarily when the Moon is closer to one of the standstills.

## Conclusion

We have used data of ancient eclipses recorded in the Indian subcontinent and compared them with the NASA calculations of ancient eclipses. We have identified nine periods of anomalous eclipses consisting of 17 eclipses with multiple observations where the NASA calculations suggest that they should not have been visible in India. We have studied the fluctuations in the Earth–Moon system based on minor (of the order of $10^{-8}$) fluctuations due to the fact that the water mass responds to the Moon with physical displacement of the mass compared to the land mass which does not get displaced. The simulations suggest a small but significant effect of this movement, with the Earth moving faster when the Moon is at standstill. A large fraction of anomalous eclipses in fact occurs when the Moon is close to the standstill.

However, fluctuations in the rotation of the Earth cannot satisfy all anomalous eclipses, as the above discussed fluctuations can only result in small drifts in horizontal shifts in eclipse maps ($\Delta T$ corrections). However, for examples where huge spatial and time corrections are required, we need to couple the above-studied phenomenon with another mechanism(s) to account for anomalies. The possible corrections would be to look into changes in

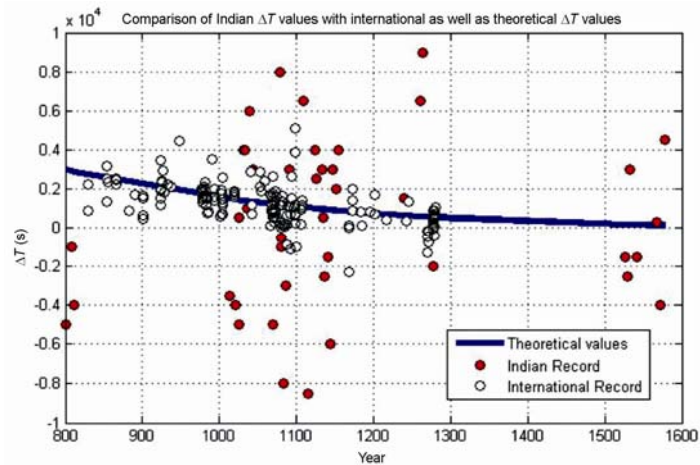

**Figure 12.** Comparison of Δ*T* values computed on the basis of the algorithm used by Soma and Tanikawa and data from other parts of the world.

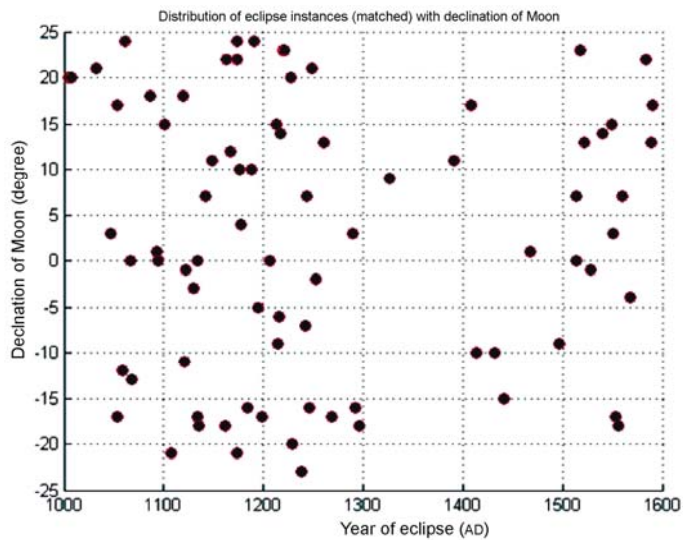

**Figure 13.** Location of the Moon at the time of eclipses are according to expectations.

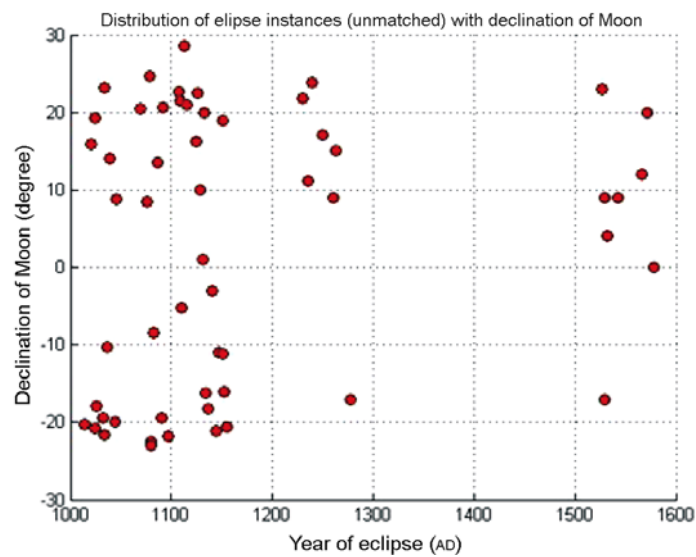

**Figure 14.** Location of the Moon at the time when the eclipses should not have been visible in India, but were recorded.

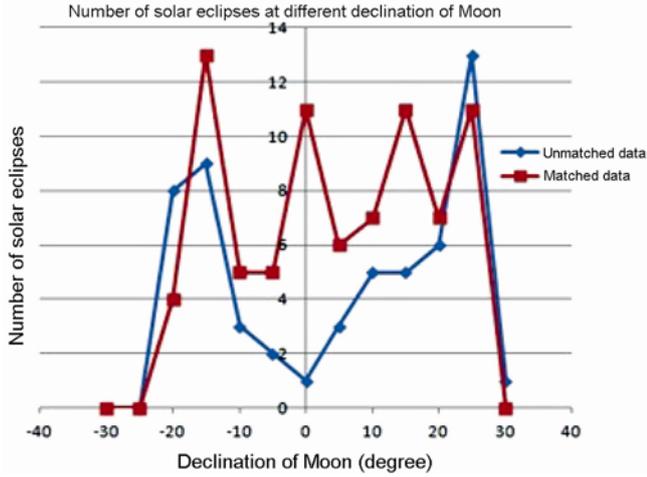

**Figure 15.** Location of the Moon during normal and anomalous eclipses. The Moon is close to its standstill position at the time of anomalous eclipses.

Moon's inclination or fluctuations in its secular acceleration at a long-term scale.

**Appendix 1.** Calculation of net force on the Moon at different declinations.

We use spherical coordinates to refer to blocks (specifically their centre of mass) as well as the Moon:

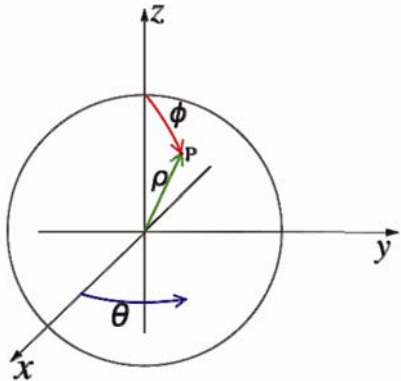

$$F_i = \frac{GMm_i}{|r_{pi} - r_m|^2} \times \frac{(r_{pi} - r_m)}{|r_{pi} - r_m|} = \frac{GMm_i}{|r_{pi} - r_m|^3}(\boldsymbol{r_{pi}} - \boldsymbol{r_m}), \quad \text{(A1.1)}$$

$$F_{net} = \sum F_i, \quad \text{(A1.2)}$$

where $r_p$ and $r_m$ are position vectors of rectangular slab and Moon respectively, given by

$$r_p = R_e \sin(\phi)\cos(\theta)i + R_e \sin(\phi)\sin(\theta)j + R_e \cos(\phi)k, \quad 0 \le \phi \le \pi, \quad 0 \le \theta \le 2\pi, \quad \text{(A1.3)}$$

$$r_m = R_m \sin(\phi)\cos(\theta)i + R_m \sin(\phi)\sin(\theta)j + R_m \cos(\phi)k, \quad 0 \le \phi \le \pi, \quad 0 \le \theta \le 2\pi. \quad \text{(A1.4)}$$

**Appendix 2.** Calculation of moment of inertia of the Earth considering its non-rigid nature.

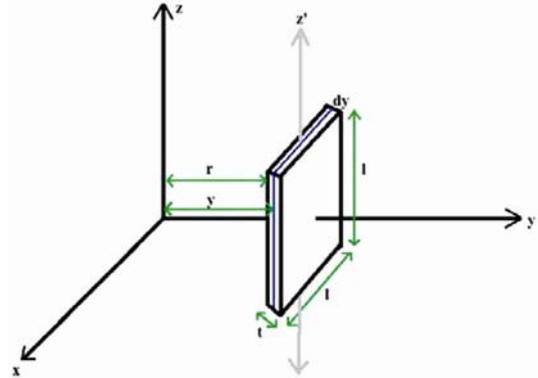

$$Iz' = \frac{l^2}{12}dm$$
$$= \frac{l^2}{12}\rho l^2 dy, \quad \text{(A2.1)}$$

$$Iz = \frac{\rho l^4}{12}dy + \rho y^2 l^2 dy$$
$$= \frac{\rho l^4}{12}\int_r^{(r+t)}dy + \rho l^2\int_r^{(r+t)}y^2 dy, \quad \text{(A2.2)}$$

$$Iz = \frac{\rho l^2 t}{12} + \frac{\rho l^2}{3[(r+t)^2 - r^2]}, \quad \text{(A2.3)}$$

$$I_{polar\,(Earth)} = \sum_{k=1}^{41482} i_k, \quad \text{(A2.4)}$$

where the parameter $r$ is variable due to influence of tides.

ACKNOWLEDGEMENTS. We thank Sir Jamsetji Tata Trust for funds. Dr M. Soma and Dr K. Tanikawa for their suggestions and assistance and valuable inputs for the study and Dr Aniket Sule and Nisha Yadav for useful discussions.

Received 7 February 2013; revised accepted 23 April 2013